\documentclass[a4paper,twocolumn,prb,floatfix,footinbib]{revtex4}

\usepackage[english]{babel}
\usepackage{latexsym}
\usepackage{graphicx}

\begin{document}
\newcommand{\uc}{{\mathrm{c}}}
\newcommand{\cs}{C$_{60}$}
\newcommand{\is}{{\em in situ}}
\newcommand{\xs}{{\em ex situ}}

\title{Single domain transport measurements of \cs\ films}

\author{S.~Rogge}
\author{M.~Durkut}
\author{T.M.~Klapwijk}
\affiliation{Department of Applied Physics and DIMES, Delft
University of Technology, Lorentzweg~1, 2628~CJ Delft, The
Netherlands}

\begin{abstract}
Thin films of potassium doped \cs, an organic semiconductor, have
been grown on silicon. The films were grown in ultra-high vacuum
by thermal evaporation of \cs\ onto oxide-terminated silicon as
well as reconstructed Si(111). The substrate termination had a
drastic influence on the \cs\ growth mode which is directly
reflected in the electrical properties of the films. Measured on
the single domain length scale, these films revealed resistivities
comparable to bulk single crystals. \textit{In situ} electrical
transport properties were correlated to the morphology of the film
determined by scanning tunneling microscopy. The observed excess
conductivity above the superconducting transition can be
attributed to two-dimensional fluctuations.
\end{abstract}

\date{\today}
\maketitle

In contrast to conventional inorganic semiconductors, the
electrical properties of organic semiconductors directly reflect
the electronic structure of the molecules they consist of. Due to
the weak Van der Waals binding between the molecules tailoring of
the electronic properties of a material is possible on the
molecular level. Presently, there is a strong interest in
high-quality thin films of organic semiconductors due to the
realization of electrostatic doping leading to metallic conduction
in these materials \cite{Jackson:98}. Intrinsic material
properties can be measured by contacting a single grain in a four
probe configuration \cite{Schoonveld:98}. This field-effect
transistor geometry is very powerful since it allows for a
continuous change in carrier density over a large range in one
sample which, is almost impossible to realize with chemical
doping.

The Van der Waals bonded organic crystals are very fragile. The
realization of high-quality thin films on a rigid substrate would
make it possible to use high resolution lithography and standard
processing, treating these materials in a similar fashion as
conventional semiconductors. This would open up the opportunity to
study the intrinsic properties of these novel materials as well as
mesoscopic physics in an unseen range of carrier densities in one
sample. As a starting point to study thin films of organic
semiconductors we have chosen the well-investigated system
K$_3$\cs. The growth as well as the superconducting transition of
this material have been studied in detail
\cite{Weaver:91,Tanigaki:92}. The motivation for starting with a
chemically instead of an electrostatically doped system is the
larger freedom in choosing a substrate. Furthermore, the
preparation of an oxide which allows for the high surface charge
necessary to achieve superconductivity with field doping is highly
nontrivial. Up to now transport studies of \cs\ films were done on
oxides in contrast to the growth studies of highly ordered films
on atomically flat substrates \cite{Weaver:91}. Here, we combine
transport and morphology studies. Both an atomically flat and an
oxide substrate are discussed: $7\times7$ reconstructed Si(111)
and SiO$_2$. Finally, for future electrostatic doping of
high-quality \cs\ films it is possible to combine atomically flat
surfaces with a gate by using Si on SiO$_2$
\cite{Weitering:00soi}.

In this paper we combine \is\ scanning tunneling microscopy (STM)
with transport experiments to link the film morphology to the
electrical properties. Fluctuation effects above the
superconducting transition were analyzed to further determine the
film quality (dimensionality). It was found that the substrate
termination has a drastic influence on the electrical properties
of the film. On oxide, ultra-thin films show thermally activated
transport. On atomically flat reconstructed Si, films remaind
superconducting down to a thickness of at least 25 monolayers
(ML). The resistivity of these films is comparable to what has
been reported for the best single crystals.

\begin{figure*}[ht]
  \centering
  \includegraphics[width=17cm]{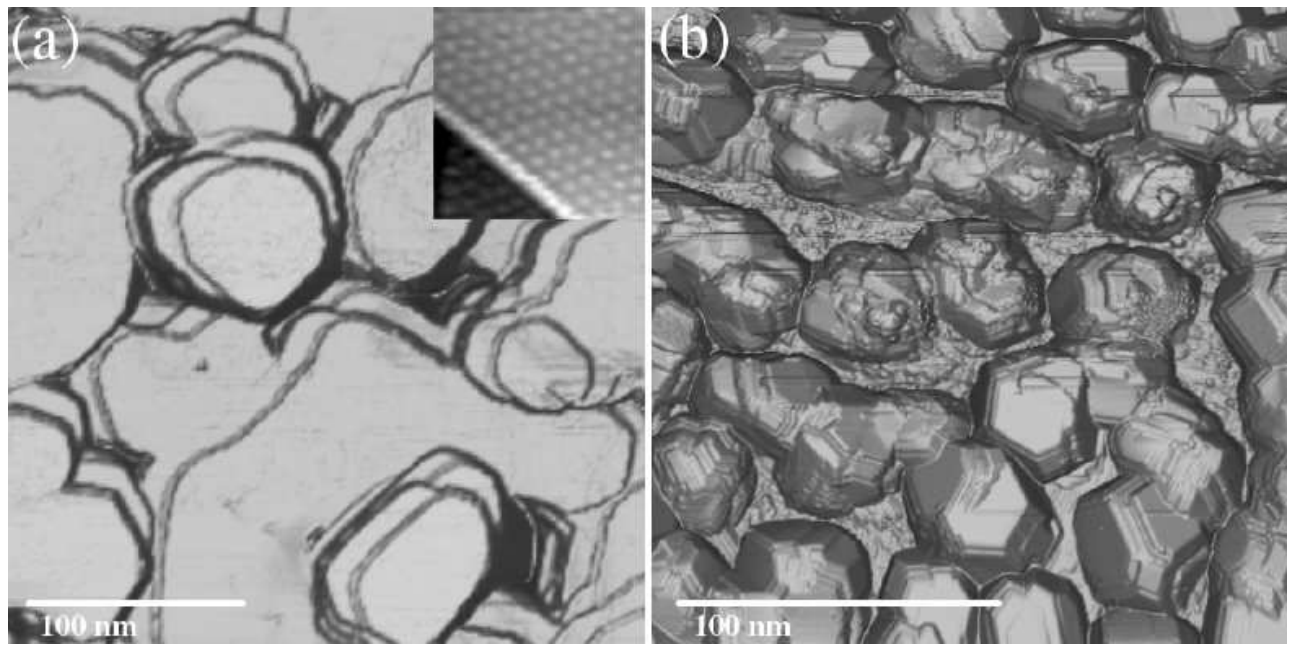}
  \caption{STM scans of K$_3$C$_{60}$ films on 7$\times$7
  terminated Si (a) and oxide terminated Si (b). The images
  represent a top view onto a pseudo three-dimensional
  light-shaded surface. The inset shows molecular ordering in
  (a). On the reconstructed substrate (a) islands/grains
  coalesce, all with the (111) surface parallel to the substrate
  (average total thickness 17~nm, height range of scan 12~nm).
  On oxide (b) the randomly oriented grains are observed (average
  total thickness 11~nm, height range 20~nm). Note the larger scan
  size of the well order film (a) compared to (b).}
  \label{fig:STM}
\end{figure*}

Film growth, STM analysis, and transport measurements were all
carried out in a single ultra-high vacuum (UHV) system with a base
pressure of $5 \cdot 10^{-11}$~mbar to prevent K oxidation.
Low-doped silicon substrates (50~$\Omega\cdot$cm and
17~k$\Omega\cdot$cm p-type wafers from standard stock) were used
to keep the background conduction low. For the films discussed
here the parallel conduction through the Si can be neglected below
225~K.

\cs\ films on two kinds of Si surface termination have been
studied, the 7$\times$7 reconstruction on Si(111) and Si oxide
which is a common substrate for transport experiments. A native
oxide was prepared by oxidizing a hydrogen-passivated Si substrate
in nitric acid. After loading into UHV the substrate was degassed
at 800~K for several hours. To prepare the 7$\times$7
reconstruction the substrate was further annealed at 1100~K for
one hour. The \cs\ was evaporated onto the room temperature
substrate from a Knudsen-cell. Films with thicknesses between 1
and 140~ML were prepared at a growth-rate of about 2~ML/hour. It
is necessary to quote mean thickness due to the Stranski-Krastanov
growth mode of \cs\ on Si leading to island formation. The
thickness was based on the \cs\ evaporation time after careful
calibration of the source based on STM scans.

Very different growth modes were observed for the two substrate
terminations. Figure~\ref{fig:STM} shows two films of similar
thickness, one on the 7$\times$7 reconstruction of Si(111) (a) and
the other on SiO$_2$ (b). The film on reconstructed Si consists of
coalesced islands/grains with the (111) surface of the \cs\ fcc
lattice parallel to the substrate as expected for the
Stranski-Krastanov growth mode. In contrast, a film of similar
thickness on an oxide-terminated surface shows small grains
without a common [111] direction. These grains show a much weaker
tendency to coalesce which manifests itself in the 2.5 times
larger roughness of the films grown on SiO$_2$.

\begin{figure}[ht]
  \centering
  \includegraphics[width=8cm]{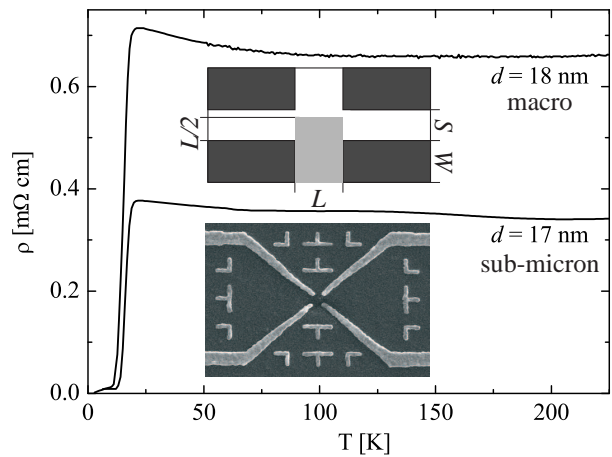}
  \caption{Films on reconstructed Si. The low resistivity trace
   was measured with the {\em sub-micron} contacts (250~nm
   probe separation, scanning electron micrograph in bottom inset).
   The top trace was measured with the {\em macro} contacts (top
   schematic, dark regions represent tungsten pads, channel length
   $L = 175$ to 600~$\mu$m, separation between current and voltage
   contacts $S = 175$~$\mu$m, and pad width $W \approx 1000$~$\mu$m).}
  \label{fig:RT}
\end{figure}

To study the electrical transport properties of the films contacts
were evaporated onto the substrate before loading it into UHV. Two
different contact arrangements were used to probe conductance at
different length scales. In one, further referred to as {\em
macro} contacts, tungsten pads were evaporated onto the substrate
by using a shadow-mask with channel lengths between 175 and
600~$\mu$m (see top inset in Fig.~\ref{fig:RT}). In the other,
further referred to as {\em sub-micron} contacts, e-beam
lithography was used to define four tungsten probes with a
separation of 250~nm (see bottom inset in Fig.~\ref{fig:RT}).

The resistance of a K$_x$C$_{60}$ film goes through a minimum as
the doping goes through the optimum value of $x=3$ ($x=0$ and
$x=6$ are both insulating with $\rho>$10$^5 \Omega\cdot$cm). This
makes it possible to follow the doping process by monitoring the
resistance. {\em Optimum doping} ({\em i.e.} $x=3$) is achieved
when the derivative of the resistance with respect to time crosses
zero. A sufficiently low K deposition-rate was used so that
diffusion times in these films could be neglected. This was
confirmed by annealing experiments which did not lead to a lower
resistivity (discussed later on in this paper).

Our ultra-thin films of optimally doped K$_3$\cs\ on
oxide-terminated Si had a considerably higher resistivity than
bulk samples and showed thermally activated transport. The sheet
resistance of the film on oxide shown in Fig.~\ref{fig:STM}b was
27~k$\Omega/\Box$ at room temperature which corresponds to a mean
resistivity of $\rho=30$~m$\Omega \cdot$cm. The slope of the
$\rho(T)$ curve is negative below 300~K as expected for a film
with such weakly linked grains. Based on transport experiments
Palstra {\em et al.} \cite{Palstra0D:92} determined a grain size
of 7~nm for their films grown on oxide which is consistent with
the morphology we observe in Fig.~\ref{fig:STM}b.

Figure~\ref{fig:RT} shows the resistivities of two films of
similar thickness on reconstructed Si measured at two different
length scales. In contrast to ultra-thin films on oxide, films of
similar thickness on reconstructed Si show a positive $\rho$ vs
$T$ slope (metallic behavior) in our measurement between 100 and
300~K. Below 100~K there is a slight rise in resistivity which can
be attributed to weak localization in these thin films. The film
measured with sub-micron contacts shows the remarkable low
resistivity of 0.35+0.14~m$\Omega \cdot$cm \footnote{Upper bound
is based on two terminal resistance for these thin high-resistive
films.} at 100~K which is at least comparable to the bulk single
crystal value of 0.5~m$\Omega \cdot$cm \cite{Hou:93}.
Table~\ref{table} lists the detailed parameters of the
measurement. The films with the lower resistivity were measured
with the sub-micron contact pattern with a probe separation of
250~nm. For the others, the macro pattern with a probe separation
of 500~$\mu$m was used. The factor 2 in resistivity may be
understood by looking at the possible origins of disorder and on
what length-scale they occur. Grain boundaries are a dominant
source of scattering in a doped semiconductor and limit the
conductivity especially considering inhomogenous doping close to
the interface. K$_3$\cs\ is a special material since it has polar
surfaces \cite{Hesper:00} which could further enhance interface
scattering. Our films on reconstructed Si consist of islands with
a typical size of 100 to 300~nm (Fig.~\ref{fig:STM}a) which is
comparable to the 250~nm probe separation of the sub-micron
pattern. Hence, direct electrical transport measurements were done
at the single domain length scale and reveal resistivities much
closer to theoretical estimates (0.12 m$\Omega\cdot$cm which are
backed by indirect measurements on bulk single crystals 0.18
m$\Omega\cdot$cm\cite{Hou:93}).

\begin{table}[ht]
  \centering
  \begin{tabular}{|c|c|c|c|c|}
    \hline
    length scale& thickness & $\rho$                & $\rho_\mathrm{upper-bound}$ & $R$ \\
    \hline
    250 nm      & 17 nm     & 0.35 m$\Omega\cdot$cm & 0.49 m$\Omega\cdot$cm & 208 $\Omega/\Box$\\
    600 $\mu$m  & 18 nm     & 0.70 m$\Omega\cdot$cm & 1.15 m$\Omega\cdot$cm & 399 $\Omega/\Box$\\
    bulk (direct)& -        & 0.50 m$\Omega\cdot$cm &                       & - \\
    bulk (indir.)& -        & 0.12 m$\Omega\cdot$cm &                       & - \\
    bulk theory  & -         & 0.18 m$\Omega\cdot$cm & 0.24 m$\Omega\cdot$cm & - \\ \hline
  \end{tabular}
  \caption{Comparison of single crystal K$_3$\cs\ properties and our
  films measured at the sub-micron and macroscopic length scale
  (upper-bounds based on the two-terminal resistance). Bulk values
  were taken from Ref.~\onlinecite{Hou:93} (indirect determination
  based on fluctuation conductivity).}
  \label{table}
\end{table}

\begin{figure}[ht]
  \centering
  \includegraphics[width=8cm]{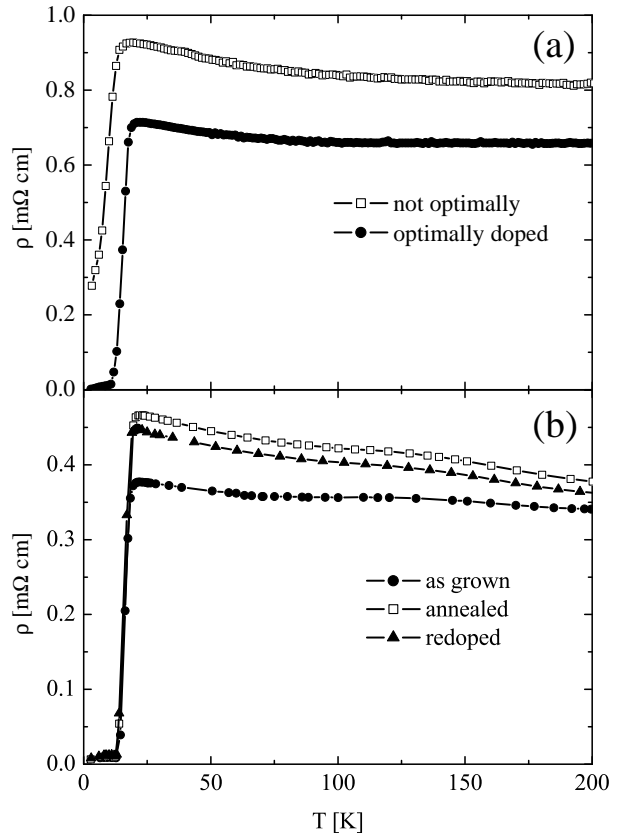}
  \caption{(a)~The graph shows two resistivity traces of a
  K$_3$\cs\ film, one optimally doped, the other after additional
  20\% K apparently leading to disorder.
  (b)~This graph shows the negative effect of annealing a film.
  The first curve is for an optimally doped film;
  the second (open symbols), after 100 minutes at 575~K;
  the third, the same film with sufficient additional K to again reach
  the lowest possible resistance at room temperature.}
  \label{fig:anneal}
\end{figure}

To study the influence of disorder we measured $\rho$ vs $T$
before and after 20\% additional potassium were supplied to an
optimally doped film, see Fig.~\ref{fig:anneal}a. The extra
potassium caused a lower critical temperature and higher
resistivity which can be attributed to disorder due to
inhomogeneous doping. To ensure that our samples are in thermal
equilibrium we did STM scans and measured $\rho(T)$ before and
after annealing of a doped film. Figure~\ref{fig:anneal}b shows a
$\rho(T)$ trace of an optimally doped film before and after
annealing at 575~K for 100 minutes. After the anneal no change in
morphology was observed in STM scans \footnote{\cs\ starts to
desorb from bulk \cs\ at 450~K. However, potassium doping elevates
this desorption temperature considerably.}; however, the
resistivity increased. Additional doping lowered the resistivity
but it remained higher than before annealing. In contrast to this,
reports on thicker films \cite{Hesper:00} and bulk samples
\cite{Xiang:92} show the lowest resistivity and sharpest
superconducting transition after annealing and re-doping cycles.
The increase in resistivity after annealing may be attributed to
the diffusion of contaminants deactivating some of the potassium.
We conclude, that our doping rate is slow enough for these
ultra-thin films and that annealing is not necessary.

\begin{figure}[ht]
  \centering
  \includegraphics[width=8cm]{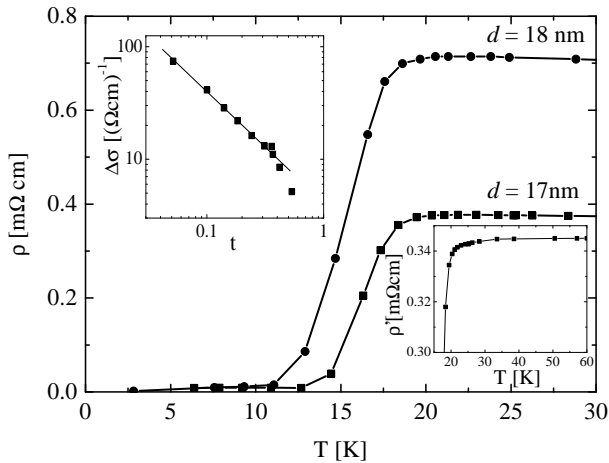}
  \caption{Superconducting transition of the same films as shown
  in Fig.~\ref{fig:RT}. The insets show the for weak localization
  corrected $\rho'(T)$ and the excess conductivity $\Delta\sigma$
  vs reduced temperature for the 17~nm film including the fit to
  the two-dimensional fluctuation model.}
  \label{fig:trans}
\end{figure}

The superconducting transition of our films is broadened as shown
in Fig.~\ref{fig:trans}. This broadening can be attributed to
fluctuation effects that lead to an enhanced conductivity above
the superconducting transition. The excess conductivity can be
described \cite{Skocpol:75} as $\Delta \sigma = \sigma_0 \cdot
t^{(d-4)/2}$. Here $t = (T-T_{\uc})/T_{\uc}$ is the reduced
temperature based on the film transition temperature and $d$ the
dimensionality of the film. To find $\Delta \sigma$, the
logarithmic contribution to $\rho(T)$ (due to weak localization)
was fitted down to 40~K and subtracted from the data. The model
for $\Delta \sigma$ with a dimensionality of two yielded a
satisfactory fit for the 17~nm ($T_{\uc}=18.5$~K) and 18~nm
($T_{\uc}=19$~K) films. As shown in the inset of
Fig.~\ref{fig:trans} we observe the two dimensional fluctuations
up to about $t=0.4$ which notably is above the bulk $T_{\uc}$. A
97~nm thick film could be fit with the three dimensional model
($T_{\uc}=17.5$~K). This is consistent with films which are not
dominated by small grains leading to zero dimensional fluctuations
independent of film thickness. The length scale is the coherence
length. However, this is only a weak indication since
magneto-resistance measurements are necessary to conclusively
study the two-dimensional effects.

Hesper {\em et al.\ }\cite{Hesper:00} reported bulk-like
resistivities (0.5~m$\Omega\cdot$cm) of 200~nm thick K$_3$\cs\
films on aluminum oxide. It seems remarkable that thin films can
achieve such good electrical properties. However, due to the Van
der Waals bonds of organic semiconductors many of the common
problems like surface states and lattice matching to the substrate
are not present. Even so the lattice mismatch between Si and
K$_3$\cs\ is $\approx 250$\% only the first layer is influenced.
The second layer is already perfectly ordered \cite{Weaver:91}.
Bulk crystals are doped after growth by intercalation
\cite{Xiang:92} of K causing inhomogenous strain due to the
lattice expansion of 0.6\%. In these films the strain can be
accommodated without creating defects in contrast to bulk samples.
Ultra-thin films of organic semiconductors may actually be easier
to grow than thicker films since strain and dopant diffusion are
less severe problems.

In conclusion, the electrical transport properties of ultra-thin
K$_3$\cs\ films have been evaluated and linked to their
morphology. Films grown on reconstructed Si showed resistivities
comparable to high-quality single crystals when measured on the
single domain length scale. These films remained superconducting
down to a mean thickness of at least 17~nm and showed two
dimensional excess conductivity due to fluctuations. Similar thin
films grown on oxide-terminated silicon had a significantly higher
resistivity and showed thermally activated transport without a
superconducting transition. The presented results are encouraging
in regard to the preparation of thin films of other organic
semiconductors since we were able to achieve bulk-like electrical
properties by simple thermal evaporation onto a suitable
substrate. This technique should be transferable to other
molecules that show highly ordered growth on Si, {\em e.g.} the
Phthalocyanines \cite{Nakamura:01}. The next step is combining
well controlled \cs\ films with electrostatic doping since
disordered films only yielded resistivities in excess of
$10^7$~$\Omega\cdot$cm \cite{Haddon:95}.

We wish to thank J.~Caro and G.D.J.~Smit for detailed discussions
concerning this work. This work is part of the research program of
the "Stichting voor Fundamenteel Onderzoek der Materie (FOM)",
which is financially supported by the "Nederlandse Organisatie
voor Wetenschappelijk Onderzoek (NWO)". One of us, S.R., wishes to
acknowledge fellowship support from the Royal Netherlands Academy
of Arts and Sciences.



\begin{thebibliography}{12}
\expandafter\ifx\csname
natexlab\endcsname\relax\def\natexlab#1{#1}\fi
\expandafter\ifx\csname bibnamefont\endcsname\relax
  \def\bibnamefont#1{#1}\fi
\expandafter\ifx\csname bibfnamefont\endcsname\relax
  \def\bibfnamefont#1{#1}\fi
\expandafter\ifx\csname citenamefont\endcsname\relax
  \def\citenamefont#1{#1}\fi
\expandafter\ifx\csname url\endcsname\relax
  \def\url#1{\texttt{#1}}\fi
\expandafter\ifx\csname
urlprefix\endcsname\relax\def\urlprefix{URL }\fi
\providecommand{\bibinfo}[2]{#2}
\providecommand{\eprint}[2][]{\url{#2}}

\bibitem[{\citenamefont{Nelson et~al.}(1998)\citenamefont{Nelson, Lin,
  Gundlach, and Jackson}}]{Jackson:98}
\bibinfo{author}{\bibfnamefont{S.~F.} \bibnamefont{Nelson}},
  \bibinfo{author}{\bibfnamefont{Y.~Y.} \bibnamefont{Lin}},
  \bibinfo{author}{\bibfnamefont{D.~J.} \bibnamefont{Gundlach}},
  \bibnamefont{and} \bibinfo{author}{\bibfnamefont{T.~N.}
  \bibnamefont{Jackson}}, \bibinfo{journal}{Appl. Phys. Lett.}
  \textbf{\bibinfo{volume}{72}}, \bibinfo{pages}{1854} (\bibinfo{year}{1998}).

\bibitem[{\citenamefont{Schoonveld et~al.}(1998)\citenamefont{Schoonveld,
  Vrijmoeth, and Klapwijk}}]{Schoonveld:98}
\bibinfo{author}{\bibfnamefont{W.~A.} \bibnamefont{Schoonveld}},
  \bibinfo{author}{\bibfnamefont{J.}~\bibnamefont{Vrijmoeth}},
  \bibnamefont{and} \bibinfo{author}{\bibfnamefont{T.~M.}
  \bibnamefont{Klapwijk}}, \bibinfo{journal}{Appl. Phys. Lett.}
  \textbf{\bibinfo{volume}{73}}, \bibinfo{pages}{3884} (\bibinfo{year}{1998}).

\bibitem[{\citenamefont{Li et~al.}(1991)\citenamefont{Li, Chander, Patrin,
  Weaver, Chibante, and Smalley}}]{Weaver:91}
\bibinfo{author}{\bibfnamefont{Y.~Z.} \bibnamefont{Li}},
  \bibinfo{author}{\bibfnamefont{M.}~\bibnamefont{Chander}},
  \bibinfo{author}{\bibfnamefont{J.}~\bibnamefont{Patrin}},
  \bibinfo{author}{\bibfnamefont{J.}~\bibnamefont{Weaver}},
  \bibinfo{author}{\bibfnamefont{L.~P.~F.} \bibnamefont{Chibante}},
  \bibnamefont{and} \bibinfo{author}{\bibfnamefont{R.}~\bibnamefont{Smalley}},
  \bibinfo{journal}{Science} \textbf{\bibinfo{volume}{253}},
  \bibinfo{pages}{429} (\bibinfo{year}{1991}).

\bibitem[{\citenamefont{Tanigaki et~al.}(1992)\citenamefont{Tanigaki, Hirosawa,
  Ebbesen, Mizuki, Shimakawa, Kubo, Tsai, and Kuroshima}}]{Tanigaki:92}
\bibinfo{author}{\bibfnamefont{K.}~\bibnamefont{Tanigaki}},
  \bibinfo{author}{\bibfnamefont{I.}~\bibnamefont{Hirosawa}},
  \bibinfo{author}{\bibfnamefont{T.~W.} \bibnamefont{Ebbesen}},
  \bibinfo{author}{\bibfnamefont{J.}~\bibnamefont{Mizuki}},
  \bibinfo{author}{\bibfnamefont{Y.}~\bibnamefont{Shimakawa}},
  \bibinfo{author}{\bibfnamefont{Y.}~\bibnamefont{Kubo}},
  \bibinfo{author}{\bibfnamefont{J.~S.} \bibnamefont{Tsai}}, \bibnamefont{and}
  \bibinfo{author}{\bibfnamefont{S.}~\bibnamefont{Kuroshima}},
  \bibinfo{journal}{Nature} \textbf{\bibinfo{volume}{356}},
  \bibinfo{pages}{419} (\bibinfo{year}{1992}).

\bibitem[{\citenamefont{Noh et~al.}(2000)\citenamefont{Noh, Jellison, Namavar,
  and Weitering}}]{Weitering:00soi}
\bibinfo{author}{\bibfnamefont{M.}~\bibnamefont{Noh}},
  \bibinfo{author}{\bibfnamefont{G.~E.} \bibnamefont{Jellison}},
  \bibinfo{author}{\bibfnamefont{F.}~\bibnamefont{Namavar}}, \bibnamefont{and}
  \bibinfo{author}{\bibfnamefont{H.~H.} \bibnamefont{Weitering}},
  \bibinfo{journal}{Appl. Phys. Lett.} \textbf{\bibinfo{volume}{76}},
  \bibinfo{pages}{733} (\bibinfo{year}{2000}).

\bibitem[{\citenamefont{Palstra et~al.}(1992)\citenamefont{Palstra, Haddon,
  Hebard, and Zaanen}}]{Palstra0D:92}
\bibinfo{author}{\bibfnamefont{T.~T.~M.} \bibnamefont{Palstra}},
  \bibinfo{author}{\bibfnamefont{R.~C.} \bibnamefont{Haddon}},
  \bibinfo{author}{\bibfnamefont{A.~F.} \bibnamefont{Hebard}},
  \bibnamefont{and} \bibinfo{author}{\bibfnamefont{J.}~\bibnamefont{Zaanen}},
  \bibinfo{journal}{Phys. Rev. Lett.} \textbf{\bibinfo{volume}{68}},
  \bibinfo{pages}{1054} (\bibinfo{year}{1992}).

\bibitem[{\citenamefont{Hou et~al.}(1993)\citenamefont{Hou, Crespi, Xiang,
  Vareka, Briceno, Zettl, and Cohen}}]{Hou:93}
\bibinfo{author}{\bibfnamefont{J.~G.} \bibnamefont{Hou}},
  \bibinfo{author}{\bibfnamefont{V.~H.} \bibnamefont{Crespi}},
  \bibinfo{author}{\bibfnamefont{X.-D.} \bibnamefont{Xiang}},
  \bibinfo{author}{\bibfnamefont{W.~A.} \bibnamefont{Vareka}},
  \bibinfo{author}{\bibfnamefont{G.}~\bibnamefont{Briceno}},
  \bibinfo{author}{\bibfnamefont{A.}~\bibnamefont{Zettl}}, \bibnamefont{and}
  \bibinfo{author}{\bibfnamefont{M.~L.} \bibnamefont{Cohen}},
  \bibinfo{journal}{Solid State Comm.} \textbf{\bibinfo{volume}{86}},
  \bibinfo{pages}{643} (\bibinfo{year}{1993}).

\bibitem[{\citenamefont{Hesper et~al.}(2000)\citenamefont{Hesper, Tjeng, Heeres,
  and Sawatzky}}]{Hesper:00}
\bibinfo{author}{\bibfnamefont{R.}~\bibnamefont{Hesper}},
  \bibinfo{author}{\bibfnamefont{L.~H.} \bibnamefont{Tjeng}},
  \bibinfo{author}{\bibfnamefont{A.}~\bibnamefont{Heeres}}, \bibnamefont{and}
  \bibinfo{author}{\bibfnamefont{G.~A.} \bibnamefont{Sawatzky}},
  \bibinfo{journal}{Phys. Rev. B} \textbf{\bibinfo{volume}{62}},
  \bibinfo{pages}{16 046} (\bibinfo{year}{2000}).

\bibitem[{\citenamefont{Xiang et~al.}(1992)\citenamefont{Xiang, Hou, Bracenou,
  Bareka, Mostovoy, Zettl, Crespi, and Cohen}}]{Xiang:92}
\bibinfo{author}{\bibfnamefont{X.-D.} \bibnamefont{Xiang}},
  \bibinfo{author}{\bibfnamefont{J.~G.} \bibnamefont{Hou}},
  \bibinfo{author}{\bibfnamefont{G.}~\bibnamefont{Bracenou}},
  \bibinfo{author}{\bibfnamefont{W.~A.} \bibnamefont{Bareka}},
  \bibinfo{author}{\bibfnamefont{R.}~\bibnamefont{Mostovoy}},
  \bibinfo{author}{\bibfnamefont{A.}~\bibnamefont{Zettl}},
  \bibinfo{author}{\bibfnamefont{V.~H.} \bibnamefont{Crespi}},
  \bibnamefont{and} \bibinfo{author}{\bibfnamefont{M.~L.} \bibnamefont{Cohen}},
  \bibinfo{journal}{Science} \textbf{\bibinfo{volume}{256}},
  \bibinfo{pages}{1190} (\bibinfo{year}{1992}).

\bibitem[{\citenamefont{Skocpol and Tinkham}(1975)}]{Skocpol:75}
\bibinfo{author}{\bibfnamefont{W.~J.} \bibnamefont{Skocpol}} \bibnamefont{and}
  \bibinfo{author}{\bibfnamefont{M.}~\bibnamefont{Tinkham}},
  \bibinfo{journal}{Rep. on Prog. in Phys.} \textbf{\bibinfo{volume}{38}},
  \bibinfo{pages}{1049} (\bibinfo{year}{1975}).

\bibitem[{\citenamefont{Nakamura et~al.}(2001)\citenamefont{Nakamura,
  Matsunobe, and Tokumoto}}]{Nakamura:01}
\bibinfo{author}{\bibfnamefont{M.}~\bibnamefont{Nakamura}},
  \bibinfo{author}{\bibfnamefont{T.}~\bibnamefont{Matsunobe}},
  \bibnamefont{and} \bibinfo{author}{\bibfnamefont{H.}~\bibnamefont{Tokumoto}},
  \bibinfo{journal}{J. Appl. Phys.} \textbf{\bibinfo{volume}{89}},
  \bibinfo{pages}{7860} (\bibinfo{year}{2001}).

\bibitem[{\citenamefont{Haddon et~al.}(1995)\citenamefont{Haddon, Perel,
  Morris, and Hebard}}]{Haddon:95}
\bibinfo{author}{\bibfnamefont{R.~C.} \bibnamefont{Haddon}},
  \bibinfo{author}{\bibfnamefont{A.~S.} \bibnamefont{Perel}},
  \bibinfo{author}{\bibfnamefont{R.~C.} \bibnamefont{Morris}},
  \bibnamefont{and} \bibinfo{author}{\bibfnamefont{A.~F.}
  \bibnamefont{Hebard}}, \bibinfo{journal}{Appl. Phys. Lett.}
  \textbf{\bibinfo{volume}{67}}, \bibinfo{pages}{121} (\bibinfo{year}{1995}).

\end{thebibliography}
\end{document}